\documentclass[pra,aps,showpacs,groupedaddress,superscriptaddress,twocolumn,toc=flat,biblatex,footinbib]{revtex4-1}
\usepackage{natbib}
\usepackage[table]{xcolor}
\usepackage[colorlinks=false]{hyperref}
\usepackage[utf8]{inputenc}
\usepackage{color}
\usepackage{bbm} 
\usepackage[english]{babel}
\usepackage{multirow}
\usepackage{amsfonts,amsmath,amssymb,stmaryrd}
\usepackage{placeins}
\usepackage{graphicx}
\usepackage{subfigure}  % use for side-by-side figures

\usepackage{epsfig}
\usepackage{mathrsfs}
\usepackage{verbatim}
\usepackage{centernot}
\usepackage{ulem}
\usepackage{longtable}
\usepackage{nicefrac}
\usepackage[framemethod=tikz]{mdframed}
\usepackage{siunitx}
\usepackage[nolist]{acronym}

\newcommand{\blue}{\color{black}}

\usepackage{soul}
\usepackage{array}

\usepackage{cancel,ifthen}
\newcommand{\cmnt}[2][NoInPuT]{\ifthenelse{\equal{#1}{NoInPuT}}{}{{\color{red}\sout{#1}}} {\color{blue} #2}}

\usepackage{bm}	% bold in math mode
\renewcommand{\vec}[1]{\bm{#1}}

\LTcapwidth=\textwidth

\begin{document}

%\normalem	% changes \emph back to normal after introducing ulem package.

\title{{\blue Ultrafast Demagnetization} through 
%Ultrafast 
{\blue Femtosecond} Generation of Non-thermal Magnons 
%in Iron
%:  \textit{Ab initio} parameterized calculations
}

\author{Markus Weißenhofer}
\email[]{markus.weissenhofer@fu-berlin.de}
 \affiliation{Department of Physics and Astronomy, Uppsala University, P. O. Box 516, S-751 20 Uppsala, Sweden}
\affiliation{Department of Physics, Freie Universit{\"a}t Berlin, Arnimallee 14, D-14195 Berlin, Germany}

\author{Peter M. Oppeneer}
\affiliation{Department of Physics and Astronomy, Uppsala University, P. O. Box 516, S-751 20 Uppsala, Sweden}


\date{\today}

\begin{abstract}
Ultrafast laser excitation of ferromagnetic metals gives rise to correlated, highly non-equilibrium dynamics of electrons, spins and lattice, which are, however, poorly described by the widely-used three-temperature model (3TM). Here, we develop a fully \textit{ab-initio} parameterized out-of-equilibrium theory based on a quantum kinetic approach--termed \textit{(N+2) temperature model}--that describes magnon occupation dynamics due to electron-magnon scattering. We apply this model to perform quantitative simulations on the ultrafast, laser-induced generation of magnons in iron and demonstrate that on these timescales the magnon distribution is non-thermal: predominantly high-energy magnons are created, while the magnon occupation close to the center of the Brillouin zone even decreases, due to a repopulation towards higher energy states via a so-far-overlooked scattering term. %Moreover, 
%We show that the 3TM can be derived from our model and compare it with our microscopic calculations. In doing so, 
We demonstrate that the simple relation between magnetization and temperature computed at equilibrium does not hold in the ultrafast regime and that the 3TM greatly overestimates the demagnetization. 
{\blue The ensuing Gilbert damping becomes strongly magnon wavevector dependent and requires a  description beyond the conventional Landau-Lifshitz-Gilbert spin dynamics.}
Our \textit{ab-initio}-parameterized calculations show that ultrafast generation of non-thermal magnons provides a sizable demagnetization within 200fs {\blue in excellent comparison with experimentally observed laser-induced demagnetizations.} {\blue Our investigation emphasizes} the importance of non-thermal magnon excitations  for the ultrafast demagnetization process.
\end{abstract}
\maketitle

\section{Introduction}
The discovery that magnetic order can be manipulated on sub-picosecond timescales by femtosecond laser pulses \cite{Beaurepaire1996,Scholl1997,Hohlfeld1997} has fueled the emergence of intensive experimental and theoretical research efforts {\blue in} the field of ultrafast magnetization dynamics.
What makes this field particularly interesting, apart from its technological potential in future memory and spintronic devices \cite{Stanciu2007,Melnikov2011}, is that many well-established physical paradigms cannot be simply transferred from the equilibrium to the ultrafast regime, due to its highly non-equilibrium nature. Relatedly, albeit more than 25 years of intense research, the underlying mechanisms of ultrafast demagnetization are still heavily debated \cite{Kirilyuk2010,Carva2017,Scheid2022}: while some works %\cite{Rhie2003,Cinchetti2006,You2018,Tengdin2018} 
\cite{Rhie2003,Cinchetti2006,Koopmans2010,Schellekens2013b,Mueller2013,Griepe2023}
lean towards longitudinal excitations -- i.e., the reduction of the magnetic moment carried by each atom due to the decrease of exchange splitting --   others \cite{Carpene2008,Schmidt2010,Carpene2015,Turgut2016,Yamamoto2019} hint at transverse spin excitations -- a reduction of the average magnetization due to the mutual tilting of the moments carried by different atoms -- as the main contribution. Non-local contributions due to superdiffusive spin currents \cite{Battiato2010,Battiato2012} are relevant in certain situations \cite{Malinowski2008,Rudolf2012,Bergeard2016,Xu2017}. However, it has become evident that they are most likely not the only mechanism of ultrafast demagnetization \cite{Schellekens2013,Turgut2013}.

Theoretical models describing ultrafast magnetization dynamics typically rely on a separation of electronic, phononic and -- if magnetization dynamics are to be considered separately -- spin degrees of freedom. Beaurepaire \textit{et al}.~\cite{Beaurepaire1996} introduced the three-temperature model (3TM) to  explain the flow of the energy transferred by the laser by assuming that each subsystem is internally in thermal equilibrium and the system can hence be described by three temperatures (for electrons, phonons and spins), together with the respective distributions (Fermi-Dirac and Bose-Einstein). However, it was pointed out in numerous investigations that the distributions are non-thermal on ultrafast timescales \cite{Sun1993,Sun1994,DelFatti1998,DelFatti2000,Guo2001,Carpene2006,Maldonado2020,Maldonado2017,Wilson2020,Ritzmann2020}. Also, the 3TM discards completely the transfer of angular momentum due to demagnetization, which, according to recent experiments \cite{Dornes2019,Tauchert2022}, appears to be primarily to the lattice. 

Transverse demagnetization is often studied using atomistic spin dynamics simulations based on the stochastic Landau-Lifshitz-Gilbert (LLG) equation  together with an extended Heisenberg model \cite{Brown1963,Kazantseva_2008,Nowak2007}, which can successfully reproduce experimentally measured demagnetization curves \cite{Zahn2021,Zahn2022}. The stochastic LLG is a Langevin-type equation with a coupling to a heat bath with given temperature via a single parameter, the Gilbert damping parameter. This parameter includes all possible contributions -- Fermi surface breathing, crystal defects, coupling to phonons, $s-d$ coupling, etc. \cite{Kambersky1970,Kambersky1976,Kunes2002,Gilbert2004,Ho2004,Steiauf2005,Faehnle2006,Gilmore2007} -- to damping and while it can in principle be obtained from \textit{ab initio} calculations, in practice it is typically taken from experimental measurements of ferromagnetic resonance (FMR) \cite{Farle1998}. On the one hand, this ensures the versatility of atomistic spin dynamics simulations, but on the other hand, it obscures the details of the underlying microscopic energy and angular momentum transfer processes - which are crucial for understanding the fundamentals of ultrafast demagnetization. For this reason, steps have been taken in recent years to explicitly consider the coupling of spins to phonons \cite{Ma2009,Perera2016,Assmann2019,Strungaru2021,Hellsvik2019,Sadhukhan2022,Mankovsky2022,Strungaru2022,Weissenhofer2022} and electrons \cite{Tveten2015,Brener2017,Barbeau2022}. Also, due to the classical nature of the commonly used stochastic LLG, the equilibrium magnon occupations calculated by it follow Rayleigh-Jeans rather than Bose-Einstein statistics, henceforth leading to the wrong temperature scaling of the magnetization \cite{Ashcroft76,Evans_2014}. {Implementation of quantum statistics in the spin-dynamics simulations can however provide the correct low-temperature scaling of the magnetization \cite{Halilov1997,Barker2019}.}

In this work, we investigate the laser-induced generation of magnons, the low energy transverse excitations of the spin system, due to electron-magnon scattering. We develop a quantum kinetic approach, which will be termed \textit{(N+2)-temperature model} [(N+2)TM], to perform quantitative simulations of the time evolution of the non-thermal magnon dynamics in bcc iron. Being based on \textit{ab initio} parameters and considering also non-thermal magnon distributions, our work goes {\blue well} beyond what has been done in Refs.~\cite{Tveten2015,Brener2017,Beens2022} and the conventional 3TM. In addition, we show that the 3TM and its relevant parameters can be obtained from our (N+2)TM and, with that, from \textit{ab initio} calculations.
{\blue Importantly, using \textit{ab initio} calculated input parameters, our quantum kinetic theory predicts a sizable and ultrafast demagnetization of iron within 200 fs, in excellent agreement with experiments \cite{Carpene2008}.}

\section{Out-of-equilibrium magnon dynamics model}
To describe the time evolution of the ultrafast non-thermal magnon occupation dynamics, we assume that their creation and annihilation is dominated by electron-magnon scattering processes. In this work, we use the $sp-d$ model \cite{Zener1951,Zener1951b} to describe such processes. The basic idea of both $s-d$ model and $sp-d$ model is the separation of electrons in localized ($d$ band) electrons and itinerant ($s$ band or $s$ and $p$ bands) electrons. The magnetic moments of the $d$ electrons make up the Heisenberg-type \cite{Heisenberg1928} magnetic moments of constant length, the small energy excitations of which are the magnons. The itinerant electrons are described within a Stoner-type model \cite{Stoner1938}. While an unambiguous identification of $sp$ and $d$ electrons as localized and itinerant is strictly speaking not possible, it has nonetheless been established in literature that these models provide a suitable framework for the description of electron-spin interaction in many phenomena relevant for spintronics, e.g.\ magnetic relaxation \cite{Mitchell1957,Heinrich1967,Tserkovnyak2004}, ultrafast demagnetization \cite{Manchon2012,Tveten2015,Brener2017,Beens2020,Beens2022,Barbeau2022,Remy2023} and spin torques \cite{Zhang2004}.

We assume local exchange between the itinerant and localized spins, {as given by the Hamiltonian} $ \hat{\mathcal{H}}_\mathrm{em} \sim  \sum_{i=1}^N     \hat{\vec{s}}^\mathrm{itin}    \cdot    \hat{\vec{S}}^\mathrm{loc}_i$, with $N$ being the number of atoms, and $\hat{\vec{s}}^\mathrm{itin}$ and $\hat{\vec{S}}^\mathrm{loc}_i$ the spin operators for itinerant ($sp$) electrons and localized ($d$) electrons at atom $i$. In second quantization and second order in magnon variables (details in {Method Section \ref{AppendixA}}), the Hamiltonian reads 
\begin{align}
    \begin{split}
    &\label{eq:sp_d_model_expansion}
    \hat{\mathcal{H}}_\mathrm{em}
    \approx
    -
    \Delta
    \sum_{\vec{k}\nu}
    \Big(
    \hat{c}^\dagger_{\vec{k}\nu\uparrow}\hat{c}_{\vec{k}\nu\uparrow}
    -
    \hat{c}^\dagger_{\vec{k}\nu\downarrow}\hat{c}_{\vec{k}\nu\downarrow}
    \Big)
    \\
    &-
    \Delta
    \sqrt{
    \frac{2}{SN}
    }
    \sum_{\vec{k}\nu\nu',\vec{q}}
    \Big(
    \hat{c}^\dagger_{\vec{k}+\vec{q}\nu\uparrow}
    \hat{c}_{\vec{k}\nu'\downarrow}
    \hat{b}^\dagger_{-\vec{q}}
    +
    \hat{c}^\dagger_{\vec{k}+\vec{q}\nu\downarrow}
    \hat{c}_{\vec{k}\nu'\uparrow}
    \hat{b}_{\vec{q}}
    \Big)
    \\
    &+
    \frac{\Delta}{SN} 
    \sum_{\vec{k}\nu\nu',\vec{q}\vec{q}'} 
    \Big(
    \hat{c}^\dagger_{\vec{k}-\vec{q}+\vec{q}'\nu\uparrow}
    \hat{c}_{\vec{k}\nu'\uparrow}
    -
    \hat{c}^\dagger_{\vec{k}-\vec{q}+\vec{q}'\nu\downarrow}
    \hat{c}_{\vec{k}\nu'\downarrow}
    \Big)
     \hat{b}^\dagger_{\vec{q}}
    \hat{b}_{\vec{q}'}.
    \end{split}
\end{align}
Here, $\Delta$ is the $sp-d$ exchange parameter,  $S$ is the absolute value of the localized spins, $\vec{k}$ and $\vec{q}$ are vectors in reciprocal space, $\hat{c}^{(\dagger)}_{\vec{k}\nu\sigma}$ is the fermionic electron annihilation (creation) operator for the itinerant electrons -- with $\nu$ being the band index and $\sigma \in  \{\uparrow,\downarrow\}$ -- and $\hat{b}^{(\dagger)}_{\vec{q}}$ is the bosonic magnon annihilation (creation) operator. The first term in Equation \eqref{eq:sp_d_model_expansion} describes the spin-splitting of the itinerant electrons due to the exchange with the localized magnetic moments, the second one the excitation (annihilation) of a magnon due to a spin flip process and the third one the spin-conserving scattering of a magnon and an electron from one state to another. It is worth noting that the second term leads to a transfer of both energy and angular momentum (i.e.,\ spin) -- since it can change the total number of magnons -- while the third term can only transfer energy. For this reason, this term was discarded earlier works \cite{Tveten2015,Brener2017,Barbeau2022}, however, our quantitative analysis reveals that the energy transferred by this term can exceed the energy transferred by the term first order in magnon operators.

We complete our Hamiltonian $\mathcal{H}=\hat{\mathcal{H}}_\mathrm{e}+\hat{\mathcal{H}}_\mathrm{m}+\hat{\mathcal{H}}_\mathrm{em}$ by considering $
    \hat{\mathcal{H}}_\mathrm{e}
    =
    \sum_{\vec{k}\nu\sigma}
    \varepsilon_{\vec{k}\nu \sigma}
    \hat{c}^\dagger_{\vec{k}\nu \sigma}\hat{c}_{\vec{k}\nu \sigma}
    $
    and
    $
    \hat{\mathcal{H}}_\mathrm{m}
    =
    \sum_{\vec{q}}
    \hbar\omega_{\vec{q}}
    \hat{b}^\dagger_{\vec{q}}
    \hat{b}_{\vec{q}}
$, with $\varepsilon_{\vec{k}\nu \sigma}=\varepsilon_{\vec{k}\nu} - 2\Delta\delta_{\sigma\uparrow}$ being the mode and spin dependent electron energies that are calculated from first-principles calculations and $\hbar\omega_{\vec{q}}$ being the magnon energies. Note that we have absorbed the term zero-th order in magnon variables in Equation \eqref{eq:sp_d_model_expansion} in the otherwise spin-independent $\hat{\mathcal{H}}_\mathrm{e}$.

Next, we use the Hamiltonian introduced above to construct a quantum kinetic approach for the description of the out-of-equilibrium dynamics of electrons and magnons. We define the rates of energy exchange between both subsystems as
\begin{align}
    \label{eq:q_kin_m}
    \dot{E}_\mathrm{m}
    &=
    \sum_{\vec{q}}
    \hbar \omega_{\vec{q}} \dot{n}_{\vec{q}}
    \\
    \label{eq:q_kin_e}
    \dot{E}_\mathrm{e}
    &=
    \sum_{\vec{k}\nu\sigma}
    \varepsilon_{\vec{k}\nu\sigma} \dot{f}_{\vec{k}\nu\sigma}
    =
    - 
    \sum_{\vec{q}}
    \hbar \omega_{\vec{q}} \dot{n}_{\vec{q}}.
\end{align}
where the dot represents temporal derivative and with the electron ($f_{\vec{k}\nu\sigma}$) and magnon ($n_{\vec{q}}$) occupation numbers. The equivalence in Equation \eqref{eq:q_kin_e} results from the conservation of total energy. The time derivatives of the occupation numbers can be calculated by applying Fermi's golden rule to the scattering Hamiltonian~\eqref{eq:sp_d_model_expansion}. To simplify the calculations, we further assume a thermal electron distribution and can hence introduce a single electronic temperature $T_\mathrm{e}$ that relates to the occupation of electronic states via the Fermi-Dirac distribution. This allows us to apply and also extend (by including terms second order in the bosonic operators) the ideas laid out in Allen's seminal work on electron-phonon interaction \cite{Allen1987} to electron-magnon scattering, yielding
$\dot{n}_{\vec{q}}
    =
    \big[
    n^\mathrm{BE}(\omega_{\vec{q}},T_\mathrm{e})
    -
    n_{\vec{q}}
    \big]
    \gamma_{\vec{q}}
    +
    \sum_{\vec{q}'}
    \big[
    (n_{\vec{q}}+1)
    n_{\vec{q}'}
    n^\mathrm{BE}(\omega_{\vec{q}}-\omega_{\vec{q}'},T_\mathrm{e})
    + 
    (\vec{q} \leftrightarrow \vec{q}')
    \big]
    \Gamma_{\vec{q}\vec{q}'}
$, with $ n^\mathrm{BE}(\omega_{\vec{q}},T_\mathrm{e})= [e^{\frac{\hbar \omega_{\vec{q}}}{k_\mathrm{B}T_\mathrm{e}}} -1]^{-1}$ being the Bose-Einstein distribution evaluated at the electron temperature.
The scattering rates are given by
\begin{align}
    \label{eq:gamma1}
    \gamma_{\vec{q}}
    &=
    \frac{4\pi\Delta^2}{ SN}
    \omega_{\vec{q}}
    I_{\uparrow\downarrow}(T_\mathrm{e})
    \sum_{\vec{k}\nu \nu'}
   \delta( {\varepsilon}_\mathrm{F} - {\varepsilon}_{\vec{k}-\vec{q}\nu \uparrow})
    \delta( {\varepsilon}_\mathrm{F} - {\varepsilon}_{\vec{k}\nu'\downarrow}),
    \\
    \begin{split}
    \label{eq:gamma2}
    \Gamma_{\vec{q}\vec{q}'}
    &=
    \frac{2\pi\Delta^2}{S^2N^2}
    ( \omega_{\vec{q}}
    -
    \omega_{\vec{q}'})
    \sum_{\sigma}
    I_{\sigma\sigma}(T_\mathrm{e})
    \\
    &\hspace{1em}\times
    \sum_{\vec{k}\nu\nu'} 
    \delta(\varepsilon_\mathrm{F}-\varepsilon_{\vec{k}-\vec{q}+\vec{q}'\nu\sigma })
    \delta(\varepsilon_\mathrm{F}-\varepsilon_{\vec{k}\nu'\sigma }),
    \end{split}
\end{align}
with $\varepsilon_\mathrm{F}$ being the Fermi energy. The functions $I_{\sigma\sigma'}(T_\mathrm{e})$ have the property $\lim_{T_\mathrm{e}\rightarrow 0} I_{\sigma\sigma'}(T_\mathrm{e}) = 1$ and account for the smearing of the Fermi-Dirac distribution at high electron temperatures, similar to what has been derived for electron-phonon scattering \cite{Maldonado2017}. The expression for $I_{\sigma\sigma'}(T_\mathrm{e})$ and details of the derivation of Equations \eqref{eq:gamma1}--\eqref{eq:gamma2} are in the {Method Section \ref{AppendixA}}.
Note that a comparison with linear spin-wave theory in the framework of the Landau-Lifshitz-Gilbert equation \cite{Lu2022} reveals that $\gamma_{\vec{q}}/\omega_{\vec{q}}=\alpha_{\vec{q}}$ can be viewed as a mode-dependent Gilbert damping parameter.

Due to the assumption that the electron occupation numbers follow the Fermi-Dirac distribution at all times, the change in electron energy is determined by the change in $T_\mathrm{e}$, i.e., $\dot{E}_\mathrm{e}=    \sum_{\vec{k}\nu\sigma}    \varepsilon_{\vec{k}\nu\sigma} (\partial f_{\vec{k}\nu\sigma}/\partial T_\mathrm{e}) \dot{T}_\mathrm{e}=C_\mathrm{e}\dot{T}_\mathrm{e}$, with the electronic heat capacity $C_\mathrm{e}=  \sum_{\vec{k}\nu\sigma}    \varepsilon_{\vec{k}\nu\sigma} (\partial f_{\vec{k}\nu\sigma}/\partial T_\mathrm{e})$. By additionally considering the absorption of a laser pulse with power $P(t)$ by the electrons and a coupling of the electrons to a phonon heat bath as in the 2TM, we finally obtain our out-of-equilibrium magnon dynamics model:
\begin{align}
    \begin{split}
    \label{eq:2+NTM_m}
    \dot{n}_{\vec{q}}
    &=
    \Big[
    n^\mathrm{BE}(\omega_{\vec{q}},T_\mathrm{e})
    -
    n_{\vec{q}}
    \Big]
    \gamma_{\vec{q}}\\
    &+
    \sum_{\vec{q}'}
    \Big[
    (n_{\vec{q}}+1)
    n_{\vec{q}'}
    n^\mathrm{BE}(\omega_{\vec{q}}-\omega_{\vec{q}'},T_\mathrm{e})
    + 
    (\vec{q} \leftrightarrow \vec{q}')
    \Big]
    \Gamma_{\vec{q}\vec{q}'},
    \end{split}
    \\
    \label{eq:2+NTM_e}
    \dot{T_\mathrm{e}}
    &=
    \frac{1}{C_\mathrm{e}}
    \Big[
    -\sum_{\vec{q}}
    \hbar \omega_{\vec{q}}
   \dot{n}_{\vec{q}}
    +
    G_\mathrm{ep}(T_\mathrm{p}-T_\mathrm{e})
    +
    P(t)
    \Big],
    \\
    \label{eq:2+NTM_p}
    \dot{T_\mathrm{p}}
    &
    =
    -
    \frac{G_\mathrm{ep}}{C_\mathrm{p}}
    (T_\mathrm{p}-T_\mathrm{e})
    .
\end{align}
Here, $T_\mathrm{p}$, $C_\mathrm{p}$ and $G_\mathrm{ep}$ are the phonon temperature and heat capacity and electron-phonon coupling constant, respectively. Note that we do not consider direct magnon-phonon coupling, which has been shown to be a reasonable approximation for $3d$ ferromagnets \cite{Zahn2021,Zahn2022}. We would like to point out that the non-thermal {magnon} occupations $n_{\vec{q}}$ can be translated to mode-specific temperatures via the Bose-Einstein distribution, $T_{\vec{q}}   := \hbar \omega_{\vec{q}}/(k_\mathrm{B}\ln(n_{\vec{q}}^{-1} + 1))$. Based on this -- and in distinction from the 3TM --  we term the framework provided by Equations \eqref{eq:2+NTM_m}-\eqref{eq:2+NTM_p} the \textit{(N+2)-temperature model} ((N+2)TM). Below, we reveal by solving these coupled equations numerically that they provide a viable framework to describe laser-induced ultrafast magnetization dynamics and the generation of \textit{non-thermal} magnons, going beyond the well-established 3TM. 

Before doing so, we want to shortly discuss the relation between the (N+2)TM introduced here and the 3TM. Albeit their phenomenological nature, the 2TM ($T_\mathrm{e}$ and $T_\mathrm{p}$) and the 3TM ($T_\mathrm{e}$, $T_\mathrm{p}$ and $T_\mathrm{m}$) have been successfully applied to explain a plethora of phenomena \cite{Caruso2022}, perhaps most prominently by Beaurepaire \textit{et al}.\ to describe the ultrafast demagnetization of Ni \cite{Beaurepaire1996}. Allen~\cite{Allen1987} and Manchon \textit{et al}.~\cite{Manchon2012} demonstrated that the 2TM and the 3TM can be derived from a microscopic out-of-equilibrium approach similar to the one used here. By assuming instantaneous relaxation of the magnon occupation numbers to the Bose-Einstein distribution with a single magnon temperature $T_\mathrm{m}$, our (N+2)TM reduces to the 3TM (in absence of magnon-phonon coupling),
\begin{align}
    \begin{split}
    C_\mathrm{m}
      \dot{T}_\mathrm{m}
    &=
  G_\mathrm{em}
      (
    T_\mathrm{e}
    -
    T_\mathrm{m}) ,
    \\
    C_\mathrm{e}
    \dot{T_\mathrm{e}} ~
    &=
    G_\mathrm{em}
    (
    T_\mathrm{m}
    -
    T_\mathrm{e})
    +
    G_\mathrm{ep}(T_\mathrm{p}-T_\mathrm{e})
    +
    P(t) ,
    \\
    C_\mathrm{p}
    \dot{T_\mathrm{p}}
    &
    =
   G_\mathrm{ep}
    (T_\mathrm{e}-T_\mathrm{p}),
    \label{eq:3T}
    \end{split}
\end{align}
with the magnon heat capacity $C_\mathrm{m}=\sum_{\vec{q}}     C_{\vec{q}}=\sum_{\vec{q}}    \hbar \omega_{\vec{q}} (\partial n_{\vec{q}}/\partial T_\mathrm{m})$ and the electron-magnon coupling constant 
\begin{align}
    \label{eq:G_em}
    G_\mathrm{em}&=
    \sum_{\vec{q}}
     C_{\vec{q}}
    \Big[
   \gamma_{\vec{q}}
   +
     \sum_{\vec{q}'}
     \frac{k_\mathrm{B}T_\mathrm{m} }{\hbar \omega_{\vec{q}}}
    \Gamma_{\vec{q}\vec{q}'}
    \Big].
\end{align}
Details of the derivation are found in {Method Section \ref{AppendixB}.}
The above expression goes beyond what was derived in Ref.~\cite{Manchon2012} by including terms second order in magnon variables and allows us to determine the electron-magnon coupling fully based on \textit{ab initio} parameters. We would like to point out that it can be extended further by going to higher order in the magnon variables.

\section{Results} 

{\blue
\subsection{Magnon lifetimes and Gilbert damping}}

\begin{figure}
\centering
    \includegraphics{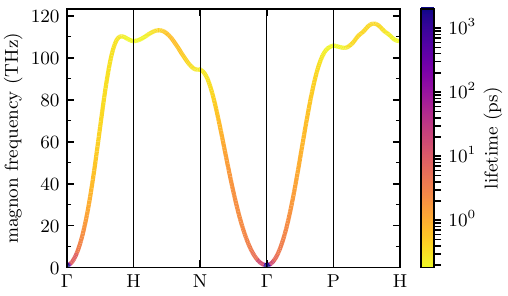}
    \caption{Magnon dispersion of bcc iron with lifetimes $\gamma_{\vec{q}}^{-1}$ {given as color code, shown along high-symmetry lines of the BZ. The lifetimes are due to the first-order contribution to the electron-magnon scattering.}}
    \label{fig:disp_1}
\end{figure}
We apply the (N+2)TM model defined by Equations \eqref{eq:2+NTM_m}-\eqref{eq:2+NTM_p} to bcc iron. To obtain a full solution of the out-of-equilibrium dynamics, it is required to calculate material specific quantities. First, we estimate $\Delta \approx \SI{0.75}{\electronvolt}$ from the band structure and with that we compute the quantities $\gamma_{\vec{q}}$, $\Gamma_{\vec{q}\vec{q}'}$ and $I_{\sigma\sigma'}(T_\mathrm{e})$, both using the full-potential linear augmented plane
wave code ELK \cite{ELK} (details can be found in the Method Section \ref{AppendixC}). For bcc iron it turns out that $I_{\sigma\sigma'}(T_\mathrm{e})$ only scales weakly with temperature and hence we use the low temperature limit $I_{\sigma\sigma'}(T_\mathrm{e})=1$ hereinafter. The parameters governing the magnon energies $\hbar\omega_{\vec{q}}= S(2 d + \sum_{j} J_{ij}[1-\exp(-i\vec{q}\cdot (\vec{r}_j-\vec{r}_i))]) $ were taken from earlier works: the exchange constants $J_{ij}$ are from first-principles calculations \cite{Mryasov1996} and the magneto-crystalline anisotropy energy $d=\SI{6.97}{\micro\electronvolt}$ per atom is from experiments \cite{Razdolski2017}. {\blue Further, we used the saturation moment $\mu_s = 2.2$\,$\mu_B$ and spin $S=2.2/2$.} Based on these parameters and the formulas derived above, we get $C_\mathrm{m}=\SI{5.720e4}{\joule\metre^{-3}\kelvin^{-1}}$ and $G_\mathrm{em}=\SI{6.796e17}{\watt\metre^{-3}\kelvin^{-1}}$ {\blue at $T_{\mathrm{m}}=\SI{300} {\,\kelvin}$}. Notably, the term first order in magnon variables leads to a contribution to $G_\mathrm{em}$ that is one order of magnitude smaller than the second-order term. We further use the room-temperature values $C_\mathrm{e}=\SI{1.013e5}{\joule\metre^{-3}\kelvin^{-1}}$, $C_\mathrm{p}=\SI{3.177e6}{\joule\metre^{-3}\kelvin^{-1}}$ and $G_\mathrm{ep}=\SI{1.051e18}{\watt\metre^{-3}\kelvin^{-1}}$ that were obtained in Refs.~\cite{Maldonado2017,Ritzmann2020} from first-principles calculations.
{\blue The influence of a temperature dependent electronic heat capacity $C_\mathrm{e}$ on the demagnetization is discussed in the Supporting Information.}

\begin{figure}[t!]
    \centering
    \includegraphics{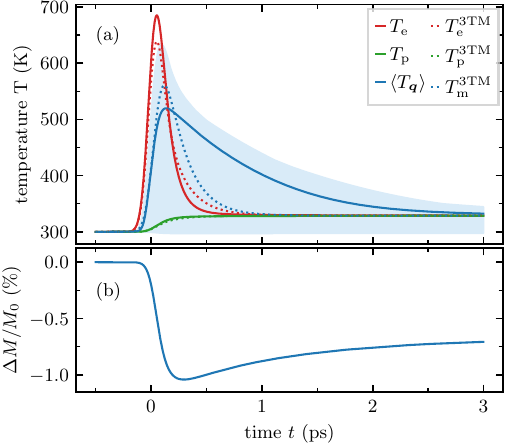}
    \includegraphics{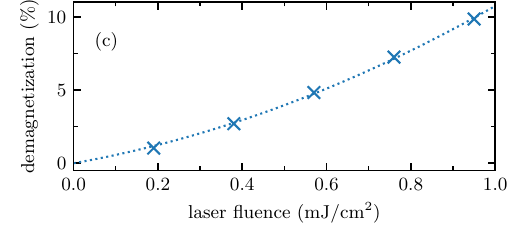}
    \caption{Laser-induced ultrafast non-equilibrium dynamics of iron calculated from an \textit{ab initio} parameterized model. (a) Temporal evolution of electron temperature $T_\mathrm{e}$, phonon temperature $T_\mathrm{p}$ and average magnon temperature $\langle T_{\vec{q}}\rangle=1/N \sum_{\vec{q}} T_{\vec{q}}$ obtained by the (N+2)TM (solid lines). The blue shaded region indicates the temperature range within which all magnon temperatures are contained. Dashed lines show the results of the 3TM solved with \textit{ab initio} calculated input parameters. (b) Relative change of total magnetization of the localized magnetic moments {$\Delta M/M_0=\sum_{\vec{q}} (n_{\vec{q}}^\mathrm{init}-n_{\vec{q}})/(NS-\sum_{\vec{q}} n_{\vec{q}}^\mathrm{init}$}), with $n_{\vec{q}}^\mathrm{init}=n^\mathrm{BE}(\omega_{\vec{q}},\SI{300}{\kelvin})$ being the occupation number before the laser pulse.
    (c)
    Demagnetization $\mathrm{max}(|\Delta M/M_0|)$ versus laser fluence computed for a ferromagnetic layer with a thickness of $\SI{20}{\nm}$. The dotted line serves as a guide to the eye.
    }
    \label{fig:2+NT_emag_1_and_2}
\end{figure}

Both $\omega_{\vec{q}}$ and the inverse of $\gamma_{\vec{q}}$, i.e.,\ the lifetime of magnons due to the contribution to electron-magnon scattering linear in the magnon variables, are shown in Figure \ref{fig:disp_1} along high-symmetry lines of the Brillouin zone (BZ). It can readily be observed that the lifetimes of high-frequency magnons {\blue are} drastically reduced as compared to the low energy ones. The {\blue $\vec{q}$-dependent} lifetimes {\blue give rise}
%relate 
to mode-specific Gilbert damping 
%parameters 
$\alpha_{\vec{q}}$ {\blue ($= \omega_{\vec{q}}/\gamma_{\vec{q}}$).} {\blue Our finding of mode-dependent Gilbert damping is consistent with experiments \cite{Li-2016} and also with a recent field-theory derivation \cite{Reyes-Osorio2023}.}   {\blue The computed $\alpha_{\vec{q}}$ values, shown in Method Section \ref{AppendixC}, range} between $\num{1.5e-3}$ and $\num{1.08e-2}$. {These values} are close to the experimentally obtained ones (via FMR measurements) for Fe ranging from $\num{1.9e-3}$ to $\num{7.2e-3}$ \cite{Oogane2006,Scheck2007,Mankovsky2013,Schoen2016,Schoen2017,Khodadadi2020}, however with a somewhat larger variation with $\vec{q}$ as compared to what was reported in Ref.~\cite{Lu2022}.

{\blue We note that the $\vec{q}$-dependent Gilbert damping goes  beyond the conventional LLG description which assumes one single damping parameter for all spin dynamics. Moreover, a further distinction between the current theory and the LLG framework is that, in the latter, there is a single damping term that governs both the energy and angular momentum transfer \cite{Ebert2011}, whereas the current theory has two terms [see Equation (\ref{eq:sp_d_model_expansion})], one that transfers energy and angular momentum and one that transfers only energy. As shown in the following, this 2$^{\rm nd}$ term is found to be important for non-thermal magnon generation.}

{\blue 
\subsection{Ultrafast dynamics}}

Based on the {\blue above-given} parameters, we calculate the coupled out-of-equilibrium magnon, electron, and phonon dynamics induced by a Gaussian laser pulse $P(t)=A/\sqrt{2\pi\zeta^2}\exp[-(t/\zeta)^2/2]$ with $A=\SI{9.619e7}{\joule\metre^{-3}}$ and $\zeta=\SI{60}{\femto\second}$ for $N=20^3$ magnon modes. Note that this value of $A$ translates to an absorbed fluence of $\SI{0.19}{\milli\joule /\cm^2}$ for a ferromagnetic layer with thickness of $\SI{20}{\nm}$, which is a typical thickness in ultrafast demagnetization experiments \cite{Beaurepaire1996}.

Figure~\ref{fig:2+NT_emag_1_and_2}(a) depicts the time evolution of electron, phonon and average magnon temperature -- together with the temperature range of all magnon temperatures -- calculated using the (N+2)TM. The electron temperature reaches a maximum of $\SI{685}{\kelvin}$ at around $\SI{52}{\femto\second}$ after the maximum of the laser pulse (located at $t=0$) and converges to the phonon temperature in less than $\SI{1.5}{\pico\second}$. The maximum of the average magnon temperature of $\SI{520}{\kelvin}$ is reached only slightly after the electronic one at around $\SI{136}{\femto\second}$, followed by a convergence to the electronic and phononic temperature to a final temperature of around $\SI{329}{\kelvin}$ at 3 ps, in agreement with what can be estimated from the energy supplied by the laser pulse and the individual heat capacities via $\Delta T=A/(C_\mathrm{m}+C_\mathrm{e}+C_\mathrm{p})=\SI{28.8}{\kelvin}$. Notably, the magnon temperatures still cover a range of around $\SI{50}{\kelvin}$ at this point in time. Our results clearly demonstrate the shortcomings of the conventional 3TM (shown as dotted lines): While the initial increase of temperatures is comparable to the (N+2)TM, {magnon thermalization} %convergence 
happens much faster in the 3TM. 
%A similar behaviour has been revealed by in Ref.~\cite{Maldonado2017}, where non-thermal phonon dynamics coupled to a electron bath were compared with the conventional 2TM.

In Figure \ref{fig:2+NT_emag_1_and_2}(b), we show the laser-induced change in magnetization (associated with the localized magnetic moments) due to the creation of additional magnons. We observe ultrafast transversal demagnetization of around one percent in less than $\SI{300}{\femto\second}$, demonstrating that the timescales obtained by our \textit{ab initio} based calculations are in reasonable agreement with experimental measurements (see, e.g.,~\cite{Carpene2008,Tengdin2018,You2018,Chekhov2021}). Notably, the minimum of the magnetization and the maximum in the average magnon temperature computed by the (N+2)TM are at different points in time. Also, the drop in the (localized) magnetization is much less pronounced than expected from the increase in average temperature: in thermal equilibrium, a temperature increase from $\SI{300}{\kelvin}$ to above $\SI{500}{\kelvin}$ approximately leads a demagnetization of $20\%$ for iron \footnote{In the classical limit (which is valid at high temperatures), the magnetization of iron scales with $M(T)/M_0=(1-T/T_\mathrm{c})^{\frac{2}{3}}$, with $T_\mathrm{c}\approx\SI{1043}{\kelvin}$ \cite{Evans_2014}.}. These observations clearly demonstrate the shortcomings of the 3TM -- where a \textit{thermal} magnon distribution at all times is assumed -- and underline the importance of treating the full, non-thermal magnon distribution in the ultrafast regime.

Figure~\ref{fig:2+NT_emag_1_and_2}(c) depicts the maximum of the demagnetization versus laser fluence for an iron layer of $\SI{20}{\nm}$. We find a nonlinear dependence, which is a result of the nonlinearity of our (N+2)TM, and a substantial demagnetization of around ten percent at $\SI{0.95}{\milli\joule \cm^{-2}}$. {\blue We note that for high fluences, higher-order magnon-magnon scattering terms that are not included in the current model could start to play a role.}
%While one could in principle go to higher fluences, we refrain from doing so, because at the current stage higher order magnon terms (i.e., magnon-magnon scattering terms) are not included in our model but could play a role for higher magnon excitation densities.
%is not suited for this, as higher order magnon terms (i.e., magnon-magnon scattering terms) are not included.}

{The obtained amount of demagnetization and the magnetization decay time (below 200 fs) for this fluence are comparable with experiments, which suggests that ultrafast magnon excitation \cite{Carpene2008,Schmidt2010,Carpene2015} provides a viable mechanism for ultrafast laser-induced demagnetization. It is also consistent with time-resolved extreme ultraviolet magneto-optical and photoemission investigations that detected magnon excitations during ultrafast demagnetization of elemental ferromagnetic samples 
\cite{Turgut2016,Eich2017}.}

\begin{figure}[t!]
    \centering
    \includegraphics{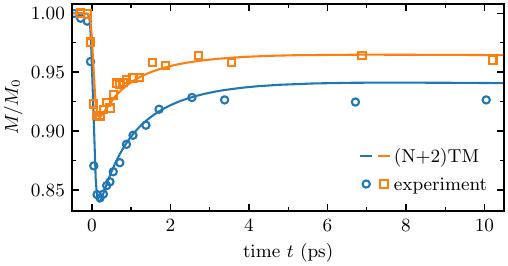}
    \caption{\blue Comparison between experiment and the (N+2)TM theory for ultrafast demagnetization in iron. The experimental data (symbols) are those of Carpene \textit{et al.}\ \cite{Carpene2008} and the solid lines are calculated from the \textit{ab initio} parameterized (N+2)TM. }
    \label{fig:comp_with_exp}
\end{figure}

{\blue For a more precise examination of the predictions of the (N+2)TM, we compare the calculated time-dependent demagnetization with experimental data for Fe in Figure  \ref{fig:comp_with_exp}. The experimental data were measured by Carpene \textit{et al.}\ \cite{Carpene2008} on a 7-nm thin film, using the time-resolved magneto-optical Kerr effect for two different pump laser fluences of $\SI{1.5}{\milli\joule\centi\metre^{-2}}$ and $\SI{3}{\milli\joule\centi\metre^{-2}}$. In the calculations we used an \textit{absorbed} laser fluence that is about five times lower, as the exact value of the absorbed fluence in experiments is difficult to estimate (due to influence of optical losses, sample reflection, etc).  Specifically, in the simulations we used absorbed laser energies of $\SI{433}{\joule\centi\metre^{-3}}$ and $\SI{693}{\joule\centi\metre^{-3}}$ in a 7-nm Fe film.
Figure \ref{fig:comp_with_exp} exemplifies that not only the amount of demagnetization but also the full 
time dependence of the demagnetization predicted by the (N+2)TM is in remarkable agreement with experiments.}\\

\begin{figure*}[ht!]
    \centering
    \includegraphics{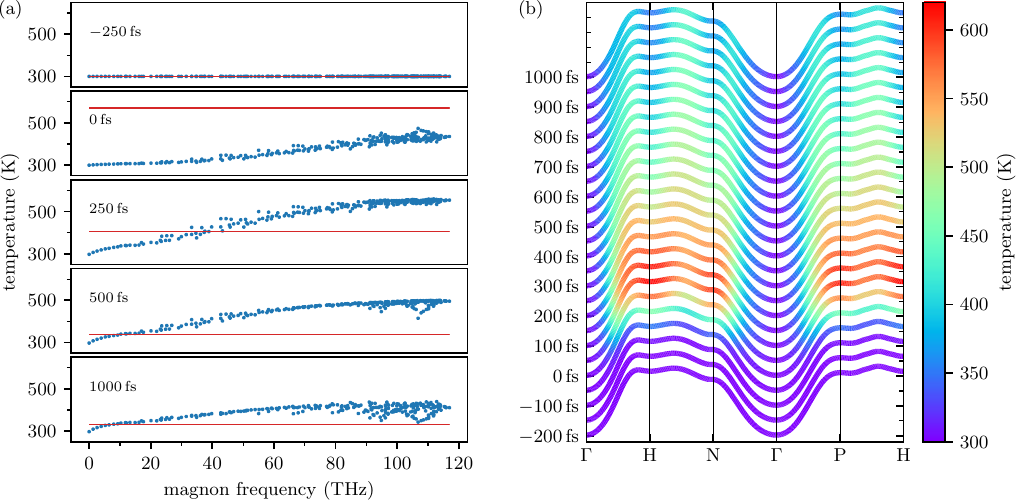}
    \caption{
    Magnon temperatures of iron during ultrafast laser excitation at different points in time (w.r.t.\ the maximum of the laser pulse) calculated from the \textit{ab initio} parameterized (N+2)TM.
    (a) Magnon temperatures (dots) versus frequency. The solid line indicates the electron temperature.
    (b) Magnon dispersion and their temperatures, depicted by the color code, shown along high-symmetry lines of the BZ.}
    \label{fig:Tq_vs_time}
\end{figure*}

{\blue
\subsection{Non-thermal magnon dynamics}}

{\blue Next, we analyze} the non-thermal magnon dynamics in more detail in Figure \ref{fig:Tq_vs_time}. There, we show the magnon temperatures versus frequency (a) and along high-symmetry lines of the BZ (b) at different points in time. The laser pulse primarily heats up high energy magnons, while the temperature of low energy magnons barely changes and even decreases slightly in the vicinity of the $\Gamma$ point (the temperatures drop by up to around $\SI{2.5}{\kelvin}$). This surprising observation is caused by a redistribution of magnons from this region to other parts of the BZ due to the term second order in the magnon operators in Equation \eqref{eq:sp_d_model_expansion}; the effective second order scattering rate $\gamma_{\vec{q}}^\mathrm{(2)}:=\sum_{\vec{q}'}\Gamma_{\vec{q}\vec{q}'}$ is negative for low magnon frequencies (more details can be found in the {Method Section \ref{AppendixC}}). It is also observed that although the magnon temperatures reached after the laser pulse are generally higher at higher frequencies, however, there is not necessarily a monotonous increase of temperature with frequency at all times: e.g., at $\SI{100}{\femto\second}$ after the laser pulse [Figure \ref{fig:Tq_vs_time}(b)], the temperatures at the points H, N, and P is higher than in between these points. Notably, the position of the maximum magnon temperature in the BZ also varies with time.\\

{\blue
\subsection{Discussion}
Different physical mechanisms have been proposed for ultrafast demagnetization in elemental $3d$ ferromagnets \cite{Rhie2003,Koopmans2010,Carpene2008,Turgut2016}. The preeminent mechanisms are Elliott-Yafet (EY) electron-phonon spin-flip scattering \cite{Koopmans2010,Mueller2013} and ultrafast magnon generation \cite{Carpene2008}. In the former, a Stoner-type picture is used to model the longitudinal reduction of the atomic moment due to electron-phonon spin-flip scattering, whereas the latter is based on length-conserving  transverse spin-wave excitations. Experimental indications of electron-phonon scattering \cite{Dornes2019,Tauchert2022} as well as of electron-magnon scattering have been reported 
\cite{Turgut2016,Eich2017}. 

The strength of the different demagnetization channels is an important issue in the on-going discussion on the dominant origin of ultrafast demagnetization \cite{Carva2017}.  \textit{Ab initio} calculated quantities such as the EY spin-flip probability are essential to achieve reliable estimates \cite{Steiauf2009,Carva2011,Carva2013}. Griepe and Atxitia \cite{Griepe2023} recently employed the microscopic 3TM \cite{Koopmans2010} and obtained quantitative agreement with measured demagnetizations for the elemental 3$d$ ferromagnets. They compared the fitted EY spin-flip probability $\alpha_{\rm sf}$  with  \textit{ab initio} calculated values \cite{Carva2013} and found these to be in good agreement, in support of an electron-phonon mechanism of ultrafast demagnetization. A drawback of their employed approach is however that only magnetization reducing spin flips are included. EY spin flips that increase the magnetization are also possible and, including these would lead to a significantly smaller demagnetization amplitude \cite{Carva2013}. This in turn would question again what amount of demagnetization is precisely due to EY electron-phonon spin flip scattering.  Conversely, in our non-thermal magnon approach we employ \textit{ab initio} calculated quantities without fit parameter. We find that the \textit{ab initio} predicted ultrafast demagnetization agrees accurately with experiments, which provides a strong support for the prominence of the non-thermal magnon channel to the ultrafast demagnetization process. 
}

\section{Conclusions}
We have developed an \textit{ab initio} parameterized quantum kinetic approach to study the laser-induced generation of magnons due to electron-magnon scattering, which we applied to iron. Our results clearly demonstrate that on ultrafast timescales the magnon distribution is non-thermal and that henceforth the simple relation between magnetization and temperature via the $M(T)$ curves computed at equilibrium does not hold: since predominantly high-energy magnons are excited the energy transferred from the laser-excited electrons creates relatively few magnons and hence the demagnetization (proportional to the total number of magnons) is much less pronounced than expected from the increase of the average magnon temperature. Notably, the number of magnons actually decreases near the center of the Brillouin zone, which is due to the scattering from low to high energy magnons by a previously neglected scattering term that can transfer energy but not angular momentum. {\blue This term, which is not included in LLG simulations, is  a crucial quantity for out-of-equilibrium magnon dynamics.}

{Our \textit{ab initio}-based calculations of the induced demagnetization in iron furthermore provide strong evidence that non-thermal magnons are excited fast and lead to a sizable demagnetization within 200 fs. {\blue The resulting time-dependent demagnetization agrees remarkably well with experiments,} which 
%in turn 
establishes the relevance of magnon excitations for the process of ultrafast optically induced demagnetization.}

\FloatBarrier

%\bibliography{bibfile.bib}
%\bibliographystyle{apsrev4-1}

%\cleardoublepage

%\cleardoublepage

\begin{widetext}
\section{Method}
\subsection{Derivation of electron-magnon scattering rates}
\label{AppendixA}

In this Method Section we derive the (N+2)TM for the description of non-thermal magnons from a microscopic Hamiltonian for electron-magnon scattering. We start with a local $sp-d$ model Hamiltonian,
\begin{align}
    \hat{\mathcal{H}}_\mathrm{em}
    =
    -
    J^\mathrm{sp-d}
    \sum_i
    \delta(\vec{r}-\vec{r}_i)
    \hat{\vec{s}}^\mathrm{itin}
    \cdot
    \vec{S}^\mathrm{loc}_i,
\end{align}
with $J^\mathrm{sp-d}$ being the $sp-d$ volume interaction energy, $\hat{\vec{s}}^\mathrm{itin}=\hat{\vec{\sigma}}$ being the spin operators of itinerant ($s$ and $p$) electrons and $\vec{S}^\mathrm{loc}_i$ being the localized ($d$) spins located at $\vec{r}_i$. For now, we treat the latter as classical vectors. The expectation value for a given spin wave function $\vec{\Psi} (\vec{r})$ is given by
\begin{align}
    \langle
    \hat{\mathcal{H}}_\mathrm{em}
    \rangle
    &=
    -
     J^\mathrm{sp-d}
    \sum_i
    \int
    \vec{\Psi}^\dagger (\vec{r})
    \delta(\vec{r}-\vec{r}_i)
    \hat{\vec{s}}^\mathrm{itin}
    \cdot
    \vec{S}^\mathrm{loc}_i
    \vec{\Psi}(\vec{r})
    {d}\vec{r}
    \\
    &=-
     J^\mathrm{sp-d}
    \sum_i
    \int
    \delta(\vec{r}-\vec{r}_i)
    \begin{pmatrix}
        \Psi^*_\uparrow (\vec{r}), & \Psi^*_\downarrow (\vec{r})
    \end{pmatrix}
    \big\{
    \hat{\sigma}_x
    S_i^x
    +
    \hat{\sigma}_y
    S_i^y
    +
    \hat{\sigma}_z
    S_i^z
    \big\}
     \begin{pmatrix}
        \Psi_\uparrow (\vec{r})\\ 
        \Psi_\downarrow (\vec{r})
    \end{pmatrix}
    {d}\vec{r}
    \\
    &=-
     J^\mathrm{sp-d}
    \sum_i
    \int
    \delta(\vec{r}-\vec{r}_i)
    \Bigg\{
    \Psi^*_\uparrow (\vec{r})\Psi_\downarrow (\vec{r})
    S_i^-
    +
    \Psi^*_\downarrow (\vec{r})\Psi_\uparrow (\vec{r})
    S_i^+
    +
    (
    \Psi^*_\uparrow (\vec{r})\Psi_\uparrow (\vec{r})
    -
    \Psi^*_\downarrow (\vec{r})\Psi_\downarrow (\vec{r})
    ) 
    S_i^z
    \Bigg\}
    {d}\vec{r}.
    \label{A4}
\end{align}
Here, we have introduced $S_i^\pm=S_i^x \pm i S_i^y$. Next, we perform a plane wave expansion of the wave functions (for a single band of itinerant electrons),
\begin{align}
    \Psi_\sigma(\vec{r})
    &=
    \frac{1}{\sqrt{V}}
    \sum_{\vec{k}}
    e^{i\vec{k}\cdot\vec{r}}
    c_{\vec{k}\sigma},
    \label{A5}
\end{align}
and a Holstein-Primakoff transformation of the localized spins,
\begin{equation}
    S_i^+=\sqrt{2S-b_i^*b_i}b_i,
    \hspace{2em}
    S_i^-=b^*_i\sqrt{2S-b_i^*b_i},
    \hspace{2em}
    S_i^z=S-b_i^*b_i,
    \label{A6}
\end{equation}
together with introducing the Fourier transform of the magnon amplitudes
\begin{equation}
   b^*_i = \frac{1}{\sqrt{N}}
   \sum_{\vec{q}}
   e^{-i \vec{q} \cdot \vec{r}_i}
   b^*_{\vec{q}},
    \hspace{2em}
    b_i = \frac{1}{\sqrt{N}}
   \sum_{\vec{q}}
   e^{i \vec{q} \cdot \vec{r}_i}
   b_{\vec{q}}.
   \label{A7}
\end{equation}
Insertion of (\ref{A5})--(\ref{A7}) into (\ref{A4}) and keeping terms up to second order in magnon variables, we get
\begin{align}
    \begin{split}
    \langle
    \hat{\mathcal{H}}_\mathrm{em}
    \rangle
     &=-
    \frac{    J^\mathrm{sp-d}}{V}
    \sum_i
    \sum_{\vec{k}\vec{k}'}
    \Bigg\{
    \sqrt{\frac{2S}{N}} 
    \sum_{\vec{q}}
    e^{-i(\vec{k}-\vec{k}'+\vec{q})\cdot\vec{r}_i}
    c^*_{\vec{k}\uparrow}
    c_{\vec{k}'\downarrow}
    b^*_{\vec{q}}
    +
    \sqrt{\frac{2S}{N}}
    \sum_{\vec{q}} 
    e^{-i(\vec{k}-\vec{k}'-\vec{q})\cdot\vec{r}_i}
    c^*_{\vec{k}\downarrow}
    c_{\vec{k}'\uparrow}
    b_{\vec{q}}
    \\
    &\hspace{6em}+
      Se^{-i(\vec{k}-\vec{k}')\cdot\vec{r}_i}
    (
    c^*_{\vec{k}\uparrow}
    c_{\vec{k}'\uparrow}
    -
    c^*_{\vec{k}\downarrow}
    c_{\vec{k}'\downarrow}
    ) - 
    \frac{1}{N} 
    \sum_{\vec{q}\vec{q}'} 
    e^{-i(\vec{k}-\vec{k}'+\vec{q}-\vec{q}')\cdot\vec{r}_i}
    (
    c^*_{\vec{k}\uparrow}
    c_{\vec{k}'\uparrow}
    -
    c^*_{\vec{k}\downarrow}
    c_{\vec{k}'\downarrow}
    )
    b^*_{\vec{q}}
    b_{\vec{q}'}
    \Bigg\}
    \end{split}
    \\
    \begin{split}
    &=
    -
    \frac{J^\mathrm{sp-d}SN}{V}
    \sum_{\vec{k}}
     (
    c^*_{\vec{k}\uparrow}
    c_{\vec{k}\uparrow}
    -
    c^*_{\vec{k}\downarrow}
    c_{\vec{k}\downarrow}
    )
    -
    \frac{J^\mathrm{sp-d}SN}{V}
    \sum_{\vec{k}\vec{q}}
    \sqrt{\frac{2}{SN}}
    \Big(
    c^*_{\vec{k}+\vec{q}\uparrow}
    c_{\vec{k}\downarrow}
    b^*_{-\vec{q}}
    +
    c^*_{\vec{k}+\vec{q}\downarrow}
    c_{\vec{k}\uparrow}
    b_{\vec{q}}
    \Big)
    \\
    &\hspace*{1.2eM} 
    +
    \frac{    J^\mathrm{sp-d}}{V}
    \sum_{\vec{k}\vec{q}\vec{q}'} 
    \Big(
    c^*_{\vec{k}-\vec{q}+\vec{q}'\uparrow}
    c_{\vec{k}\uparrow}
    -
    c^*_{\vec{k}-\vec{q}+\vec{q}'\downarrow}
    c_{\vec{k}\downarrow}
    \Big)
     b^*_{\vec{q}}
    b_{\vec{q}'}
    .
    \end{split}
\end{align}
For multiple itinerant bands and in second quantization we obtain
\begin{align}
    \begin{split}
    \hat{\mathcal{H}}_\mathrm{em}
    &=
    -
    \Delta
    \sum_{\vec{k}\nu}
     (
    \hat{c}^\dagger_{\vec{k}\nu\uparrow}
    \hat{c}_{\vec{k}\nu\uparrow}
    -
    \hat{c}^\dagger_{\vec{k}\nu\downarrow}
    \hat{c}_{\vec{k}\nu\downarrow}
    )
    -
    \Delta
    \sqrt{\frac{2}{SN}}
    \sum_{\vec{k}\nu\nu',\vec{q}}
    \Big(
    \hat{c}^\dagger_{\vec{k}+\vec{q}\nu\uparrow}
    \hat{c}_{\vec{k}\nu'\downarrow}
    \hat{b}^\dagger_{-\vec{q}}
    +
    \hat{c}^\dagger_{\vec{k}+\vec{q}\nu\downarrow}
    \hat{c}_{\vec{k}\nu'\uparrow}
    \hat{b}_{\vec{q}}
    \Big)
    \\
    &\hspace*{1.2eM} 
    +
    \frac{   \Delta}{SN}
    \sum_{\vec{k}\nu\nu',\vec{q}\vec{q}'} 
    \Big(
    \hat{c}^\dagger_{\vec{k}-\vec{q}+\vec{q}'\nu\uparrow}
    \hat{c}_{\vec{k}\nu'\uparrow}
    -
    \hat{c}^\dagger_{\vec{k}-\vec{q}+\vec{q}'\nu\downarrow}
    \hat{c}_{\vec{k}\nu'\downarrow}
    \Big)
    \hat{b}^\dagger_{\vec{q}}
    \hat{b}_{\vec{q}'}.
    \end{split}
\end{align}
where we have introduced $\Delta=\frac{J^\mathrm{sp-d}SN}{V}$. Note that due to the plane wave ansatz we have implicitly assumed that the itinerant electrons are completely delocalized and interband scattering (from $\nu$ to $\nu'\neq\nu$) fully contributes to the electron-magnon scattering.

Next, we use Fermi's golden rule to get the change of the magnon occupation number $n_{\vec{q}}=\langle\hat{b}^\dagger_{\vec{q}}\hat{b}_{\vec{q}}\rangle$. Fermi's golden rule computes the probability $W(i\rightarrow f)$ for a small perturbation term in the Hamiltonian, $\hat{H}'$ (in our specific case, $ \hat{\mathcal{H}}_\mathrm{em}$) via
\begin{equation}
W(i\rightarrow f)= \frac{2\pi}{\hbar}   |\langle f |\hat{H}'| i\rangle |^2
\delta(E_f- E_i),
\end{equation}
where  $| i\rangle$ and $|f \rangle$ denote the initial and final state, respectively.

We start with the term first order in the magnon variables,
\begin{align}
\begin{split}
    \label{eq:SM_Nq_1_1}
  \dot{n}_{\vec{q}}^\mathrm{(1)}
  &=
  W({n}_{\vec{q}}\rightarrow {n}_{\vec{q}}+1)
  -
  W({n}_{\vec{q}}\rightarrow {n}_{\vec{q}}-1)
  \\
    &=
    \frac{2\pi}{\hbar}
    \frac{2\Delta ^2}{S N}
    \sum_{\vec{k}\nu\nu'}
    \big\{
    (1-f_{\vec{k}-\vec{q}\nu\uparrow})
    f_{\vec{k}\nu'\downarrow}
    -
    (
    f_{\vec{k}-\vec{q}\nu\uparrow}
    -
    f_{\vec{k}\nu'\downarrow}
    )
    n_{\vec{q}}
    \big\}
    \delta({\varepsilon}_{\vec{k}\nu'\downarrow}- {\varepsilon}_{\vec{k}-\vec{q}\nu\uparrow} -\hbar\omega_{\vec{q}} ),
\end{split}
\end{align}
with $f_{\vec{k}\nu\sigma}=\langle\hat{c}^\dagger_{\vec{k}\nu\sigma}\hat{c}_{\vec{k}\nu\sigma}\rangle$ and $\varepsilon_{\vec{k}\nu\sigma}$ and $\hbar\omega_{\vec{q}}$ being the eigenenergies of electrons and magnons, respectively.

Hereinafter, we make the assumption that due to the fast equilibration processes for electrons, they always follow the Fermi-Dirac distribution, $
    f^\mathrm{FD}(\varepsilon_{\vec{k}\nu\sigma},T_\mathrm{e})
    =
   [
    e^{({\varepsilon}_{\vec{k}\nu\sigma}-\varepsilon_\mathrm{F})/k_\mathrm{B} T_\mathrm{e}}
    +
    1
    ]^{-1},
$
with a single electron temperature $T_\mathrm{e}$.
Before we continue we need the following relation,
\begin{align}
    \begin{split}
    f^\mathrm{FD}(\varepsilon_{\vec{k}\nu'\downarrow},T_\mathrm{e})
    (1-f^\mathrm{FD}(\varepsilon_{\vec{k}-\vec{q}\nu\uparrow},T_\mathrm{e}))&\delta({\varepsilon}_{\vec{k}\nu'\downarrow}- {\varepsilon}_{\vec{k}-\vec{q}\nu\uparrow} -\hbar\omega_{\vec{q}} ) =\\
    &
    (
    f^\mathrm{FD}(\varepsilon_{\vec{k}-\vec{q}\nu\uparrow},T_\mathrm{e})
    -
    f^\mathrm{FD}(\varepsilon_{\vec{k}\nu'\downarrow},T_\mathrm{e})
    )
    n^\mathrm{BE}(\omega_{\vec{q}},T_\mathrm{e})
    \delta({\varepsilon}_{\vec{k}\nu'\downarrow}- {\varepsilon}_{\vec{k}-\vec{q}\nu\uparrow} -\hbar\omega_{\vec{q}} )
    \end{split}
\end{align}
with $ n^\mathrm{BE}(\omega_{\vec{q}},T_\mathrm{e})= [e^{\frac{\hbar \omega_{\vec{q}}}{k_\mathrm{B}T_\mathrm{e}}} -1]^{-1}$ being the Bose-Einstein distribution evaluated at the electron temperature.
Now we can simplify Equation \eqref{eq:SM_Nq_1_1}, yielding
\begin{align}
\label{eq:SM_Nq_1_2}
  \dot{n}_{\vec{q}}^\mathrm{(1)} 
    &\approx
    \frac{2\pi}{\hbar}
    \frac{2\Delta ^2}{S N}
    \sum_{\vec{k}\nu\nu'}
    \big[
    n^\mathrm{BE}(\omega_{\vec{q}},T_\mathrm{e})
    -
    n_{\vec{q}}
    \big]
     (
    f^\mathrm{FD}(\varepsilon_{\vec{k}-\vec{q}\nu\uparrow},T_\mathrm{e})
    -
    f^\mathrm{FD}(\varepsilon_{\vec{k}\nu'\downarrow},T_\mathrm{e})
    )
    \delta({\varepsilon}_{\vec{k}\nu'\downarrow}- {\varepsilon}_{\vec{k}-\vec{q}\nu\uparrow} -\hbar\omega_{\vec{q}} )
    \nonumber \\
    &
    =~\big[
    n^\mathrm{BE}(\omega_{\vec{q}},T_\mathrm{e})
    -
    n_{\vec{q}}
    \big]
    \gamma_{\vec{q}}.
\end{align}
With $\gamma_{\vec{q}}$ being the linewidth -- i.e., the inverse lifetime -- of the magnon due to the first order contribution to electron-magnon scattering. Following the ideas laid out by Allen~\cite{Allen1987} and Maldonado \textit{et al.}~\cite{Maldonado2017}, it can be computed as
\begin{align}
    \gamma_{\vec{q}}
    &=
     \frac{2\pi}{\hbar}
    \frac{2\Delta ^2}{S N}
    \sum_{\vec{k}\nu\nu'}
    [
    f^\mathrm{FD}(\varepsilon_{\vec{k}-\vec{q}\nu\uparrow},T_\mathrm{e})
    -
    f^\mathrm{FD}(\varepsilon_{\vec{k}\nu'\downarrow},T_\mathrm{e})
    ]
   \delta({\varepsilon}_{\vec{k}\nu'\downarrow}- {\varepsilon}_{\vec{k}-\vec{q}\nu\uparrow} -\hbar\omega_{\vec{q}} )
    \\
    &=
      \frac{2\pi}{\hbar}
    \frac{2\Delta ^2}{S N}
    \sum_{\vec{k}\nu\nu'}
    \int d{\varepsilon}\
    \delta( {\varepsilon} - {\varepsilon}_{\vec{k}-\vec{q}\nu\uparrow})
    \int d{\varepsilon}'\
    \delta( {\varepsilon}' - {\varepsilon}_{\vec{k}\nu'\downarrow})
    [
    f^\mathrm{FD}({\varepsilon},T_\mathrm{e})
    -
    f^\mathrm{FD}({\varepsilon}',T_\mathrm{e})
    ]
     \delta(
    {\varepsilon}'- {\varepsilon} 
    -\hbar\omega_{\vec{q}} 
    )
   \\
    &\approx
    \frac{2\pi}{\hbar}
    \frac{2\Delta ^2}{S N}
      \sum_{\vec{k}\nu\nu'}
    \delta( {\varepsilon}_\mathrm{F} - {\varepsilon}_{\vec{k}-\vec{q}\nu \uparrow})
    \delta( {\varepsilon}_\mathrm{F} - {\varepsilon}_{\vec{k}\nu'\downarrow})
    \int d{\varepsilon}\
    \int d{\varepsilon}'\
    [
    f^\mathrm{FD}({\varepsilon},T_\mathrm{e})
    -
    f^\mathrm{FD}({\varepsilon}',T_\mathrm{e})
    ]
     \delta(
    {\varepsilon}'- {\varepsilon} 
    -\hbar\omega_{\vec{q}} 
    )
    \frac{g_\uparrow( {\varepsilon})g_\downarrow( {\varepsilon}')}{g_\uparrow({\varepsilon}_\mathrm{F}) g_\downarrow({\varepsilon}_\mathrm{F})}
    \\
    &\approx
\frac{2\pi}{\hbar}
    \frac{2\Delta ^2}{S N}
    \hbar \omega_{\vec{q}}
     \sum_{\vec{k}\nu\nu'}
    \delta( {\varepsilon}_\mathrm{F} - {\varepsilon}_{\vec{k}-\vec{q}\nu \uparrow})
    \delta( {\varepsilon}_\mathrm{F} - {\varepsilon}_{\vec{k}\nu'\downarrow})
    \int d{\varepsilon}\
    (-1)
     \frac{\partial f^\mathrm{FD}({\varepsilon},T_\mathrm{e})}{\partial {\varepsilon}}
    \frac{g_\uparrow( {\varepsilon})g_\downarrow( {\varepsilon}+\hbar \omega_{\vec{q}})}{g_\uparrow({\varepsilon}_\mathrm{F}) g_\downarrow({\varepsilon}_\mathrm{F})}
   \\
   &\approx
\frac{2\pi}{\hbar}
    \frac{2\Delta ^2}{S N}
    \hbar \omega_{\vec{q}}
    \sum_{\vec{k}\nu\nu'}
    \delta( {\varepsilon}_\mathrm{F} - {\varepsilon}_{\vec{k}-\vec{q}\nu \uparrow})
    \delta( {\varepsilon}_\mathrm{F} - {\varepsilon}_{\vec{k}\nu'\downarrow})
    \int d{\varepsilon}\
    (-1)
     \frac{\partial f^\mathrm{FD}({\varepsilon},T_\mathrm{e})}{\partial {\varepsilon}}
    \frac{g_\uparrow( {\varepsilon})g_\downarrow( {\varepsilon})}{g_\uparrow({\varepsilon}_\mathrm{F}) g_\downarrow({\varepsilon}_\mathrm{F})}
     \\
    &=
    \frac{4\pi\Delta^2}{N S}
    \omega_{\vec{q}}
   \sum_{\vec{k}\nu\nu'}
    \delta( {\varepsilon}_\mathrm{F} - {\varepsilon}_{\vec{k}-\vec{q}\nu \uparrow})
    \delta( {\varepsilon}_\mathrm{F} - {\varepsilon}_{\vec{k}\nu'\downarrow})
    I_{\uparrow\downarrow}(T_\mathrm{e})
\end{align}
with $\varepsilon_\mathrm{F}$ being the Fermi energy, the spin-dependent density of states is $g_\sigma(\varepsilon) = \sum_{\vec{k}\nu} \delta(\varepsilon-\varepsilon_{\vec{k}\nu\sigma})$ and the thermal correction factor given by
\begin{equation}
    I_{\sigma\sigma'}(T_\mathrm{e})
    =
    \int d{\varepsilon}\
    (-1)
    \frac{\partial f^\mathrm{FD}({\varepsilon},T_\mathrm{e})}{\partial {\varepsilon}}
     \frac{g_\sigma( {\varepsilon})g_\sigma'( {\varepsilon})}{g_\sigma({\varepsilon}_\mathrm{F}) g_\sigma'({\varepsilon}_\mathrm{F})}.
\end{equation}
 It is obvious that $\lim_{T_\mathrm{e}\rightarrow 0} I_{\sigma\sigma'}(T_\mathrm{e}) = 1$. Note that we have used that the energy scale of magnons is much smaller than the one of electrons, i.e., that $\hbar\omega_{\vec{q}}\ll \varepsilon,\varepsilon'$.

The contribution of the term second order in magnon variables to the occupation number can be calculated analogous and reads 
\begin{align}
    \begin{split}
    &\dot{n}_{\vec{q}}^\mathrm{(2)}
     =
     \frac{2\pi}{\hbar}
     \Big(
    \frac{   \Delta}{SN}
    \Big)^2
    \sum_{\vec{k}\nu\nu'\sigma,\vec{q}'} 
    \Big\{
    ({n}_{\vec{q}}+1)
    {n}_{\vec{q}'}
     \Big(
    (1-f^\mathrm{FD}(\varepsilon_{\vec{k}-\vec{q}+\vec{q}'\nu\sigma},T_\mathrm{e}))
    f^\mathrm{FD}(\varepsilon_{\vec{k}\nu'\sigma},T_\mathrm{e})
    \delta(\hbar\omega_{\vec{q}}-\hbar\omega_{\vec{q}'}+\varepsilon_{\vec{k}-\vec{q}+\vec{q}'\nu\sigma}-\varepsilon_{\vec{k}\nu'\sigma})
    \Big)
     \\
    &\hspace{5em}
    -
    \Big(
    \vec{q}\leftrightarrow\vec{q}'
    \Big)
    \Big\}
    \end{split}
    \\
    \begin{split}
    &=
   \frac{2\pi}{\hbar}
     \Big(
    \frac{   \Delta}{SN}
    \Big)^2 \!
    \sum_{\vec{k}\nu\nu'\sigma,\vec{q}'} \!
    \Big\{
     ({n}_{\vec{q}}+1)
    {n}_{\vec{q}'}
     n^\mathrm{BE}(\omega_{\vec{q}}-\omega_{\vec{q}'},T_\mathrm{e})
    \Big(    
    f^\mathrm{FD}(\varepsilon_{\vec{k}-\vec{q}+\vec{q}'\nu\sigma},T_\mathrm{e})
    -
   f^\mathrm{FD}(\varepsilon_{\vec{k}\nu'\sigma},T_\mathrm{e})
     \Big) \times \\
      &\hspace{7em}
    \delta(\hbar\omega_{\vec{q}}-\hbar\omega_{\vec{q}'}+\varepsilon_{\vec{k}-\vec{q}+\vec{q}'\nu\sigma}-\varepsilon_{\vec{k}\nu'\sigma})
    -
    \Big(
    \vec{q}\leftrightarrow\vec{q}'
    \Big)
    \Big\}
    \end{split}
    \\
    \begin{split}
    &\approx
   \frac{2\pi}{\hbar}
     \Big(
    \frac{   \Delta}{SN}
    \Big)^2 \!
    \sum_{\vec{k}\nu\nu'\sigma,\vec{q}'} \!
    \Big\{
     ({n}_{\vec{q}}+1)
    {n}_{\vec{q}'}
    n^\mathrm{BE}(\omega_{\vec{q}}-\omega_{\vec{q}'},T_\mathrm{e})
   (\hbar \omega_{\vec{q}}-\hbar\omega_{\vec{q}'})
   \delta(\varepsilon_\mathrm{F}-\varepsilon_{\vec{k}-\vec{q}+\vec{q}'\nu\sigma})
   \delta(\varepsilon_\mathrm{F}-\varepsilon_{\vec{k}\nu'\sigma})
   I_{\sigma\sigma}(T_\mathrm{e})
    -
    \Big(
    \vec{q}\leftrightarrow\vec{q}'
    \Big)
    \Big\}
    \end{split}
    \\
  \begin{split}
    &=
   \frac{2\pi}{\hbar}
     \Big(
    \frac{   \Delta}{SN}
    \Big)^2
    \sum_{\vec{q}'} 
    \Big\{
     ({n}_{\vec{q}}+1)
    {n}_{\vec{q}'}
    n^\mathrm{BE}(\omega_{\vec{q}}-\omega_{\vec{q}'},T_\mathrm{e})
    +
    \big(
    \vec{q}\leftrightarrow\vec{q}'
    \big)
    \Big\} \!\!\!
    \sum_{\vec{k}\nu\nu'\sigma} 
   (\hbar \omega_{\vec{q}}-\hbar\omega_{\vec{q}'})
   \delta(\varepsilon_\mathrm{F}-\varepsilon_{\vec{k}-\vec{q}+\vec{q}'\nu\sigma})
   \delta(\varepsilon_\mathrm{F}-\varepsilon_{\vec{k}\nu'\sigma})
   I_{\sigma\sigma}(T_\mathrm{e})
    \end{split}
      \\
  \begin{split}
    &=
   \frac{2\pi}{\hbar}
     \Big(
    \frac{   \Delta}{SN}
    \Big)^2
    \sum_{\vec{q}'} 
    \Big\{
     ({n}_{\vec{q}}+1)
    {n}_{\vec{q}'}
    n^\mathrm{BE}(\omega_{\vec{q}}-\omega_{\vec{q}'},T_\mathrm{e})
    +
    \big(
    \vec{q}\leftrightarrow\vec{q}'
    \big)
    \Big\} \!\!\!
    \sum_{\vec{k}\nu\nu'\sigma} 
   (\hbar \omega_{\vec{q}}-\hbar\omega_{\vec{q}'})
   \delta(\varepsilon_\mathrm{F}-\varepsilon_{\vec{k}-\vec{q}+\vec{q}'\nu\sigma})
   \delta(\varepsilon_\mathrm{F}-\varepsilon_{\vec{k}\nu'\sigma})
   I_{\sigma\sigma}(T_\mathrm{e})
    \end{split}  \\
  \begin{split}
    &=
    \sum_{\vec{q}'} 
    \Big\{
     ({n}_{\vec{q}}+1)
    {n}_{\vec{q}'}
    n^\mathrm{BE}(\omega_{\vec{q}}-\omega_{\vec{q}'},T_\mathrm{e})
    +
    \big(
    \vec{q}\leftrightarrow\vec{q}'
    \big)
    \Big\}
    \Gamma_{\vec{q}\vec{q}'}(T_\mathrm{e})
    \end{split}
\end{align}
with
\begin{equation}
    \Gamma_{\vec{q}\vec{q}'}(T_\mathrm{e})
    =
      \frac{2\pi}{\hbar}
     \Big(
    \frac{   \Delta}{SN}
    \Big)^2
     (\hbar \omega_{\vec{q}}-\hbar\omega_{\vec{q}'})
     \sum_{\sigma} 
     I_{\sigma\sigma}(T_\mathrm{e})
     \sum_{\vec{k}\nu\nu'} 
   \delta(\varepsilon_\mathrm{F}-\varepsilon_{\vec{k}-\vec{q}+\vec{q}'\nu\sigma})
   \delta(\varepsilon_\mathrm{F}-\varepsilon_{\vec{k}\nu'\sigma}).
\end{equation}

%\newpage

\subsection{Derivation of the three temperature model}
\label{AppendixB}

In what follows, it is demonstrated that the three temperature model (3TM) can be obtained from the (N+2)-temperature model derived in the main text,
\begin{align}
    \begin{split}
    \dot{n}_{\vec{q}}
    &=
    \Big[
    n^\mathrm{BE}(\omega_{\vec{q}},T_\mathrm{e})
    -
    n_{\vec{q}}
    \Big]
    \gamma_{\vec{q}}+
    \sum_{\vec{q}'}
    \Big[
    (n_{\vec{q}}+1)
    n_{\vec{q}'}
    n^\mathrm{BE}(\omega_{\vec{q}}-\omega_{\vec{q}'},T_\mathrm{e})
    + 
    (\vec{q} \leftrightarrow \vec{q}')
    \Big]
    \Gamma_{\vec{q}\vec{q}'},
    \end{split}
    \\
    \dot{T_\mathrm{e}}
    &=
    \frac{1}{C_\mathrm{e}}
    \Big[
    -\sum_{\vec{q}}
    \hbar \omega_{\vec{q}}
   \dot{n}_{\vec{q}}
    +
    G_\mathrm{ep}(T_\mathrm{p}-T_\mathrm{e})
    +
    P(t)
    \Big] ,
    \\
    \dot{T_\mathrm{p}}
    &
    =
    -
    \frac{G_\mathrm{ep}}{C_\mathrm{p}}
    (T_\mathrm{p}-T_\mathrm{e}),    
\end{align}
by assuming instantaneous relaxation of the magnon occupation numbers to the Bose-Einstein distribution with a single magnon temperature $T_\mathrm{m}$, i.e., $n_{\vec{q}}=n^\mathrm{BE}(\omega_{\vec{q}},T_\mathrm{m})$. For the sake of readability we rewrite $n^\mathrm{BE}(\omega_{\vec{q}},T_\mathrm{m})=n_{\vec{q}}(T_\mathrm{m})$.

We start with the first order scattering term:
\begin{align}
    \dot{n}_{\vec{q}}^\mathrm{(1)}
    &=
    [
    n_{\vec{q}}(T_\mathrm{e})
    -
    n_{\vec{q}}(T_\mathrm{m})
    ]
    \gamma_{\vec{q}}
    \approx
    (T_\mathrm{e}
    -
    T_\mathrm{m})
    \frac{\partial n_{\vec{q}}(T)}{ \partial T}\bigg|_{T=T_\mathrm{m}}
    \gamma_{\vec{q}}(T_\mathrm{e})
    =
     (
    T_\mathrm{e}
    -
    T_\mathrm{m})
    \frac{C_{\vec{q}}\gamma_{\vec{q}}}{\hbar \omega_{\vec{q}}}.
\end{align}
Here we have introduced the mode-dependent magnon heat capacity $C_{\vec{q}}=\hbar \omega_{\vec{q}} \frac{\partial n_{\vec{q}}(T_\mathrm{m})}{ \partial T}$.

In order to calculate the scattering term second order in the magnon variables, we first introduce the following relation 
\begin{align}
    \big(n_{\vec{q}'}(T_\mathrm{m})+1\big)n_{\vec{q}}(T_\mathrm{m})
    =
    \big[n_{\vec{q}'}(T_\mathrm{m}) -n_{\vec{q}}(T_\mathrm{m})\big]
    n_{\vec{q}-\vec{q}'}(T_\mathrm{m}).
\end{align}
Now we calculate
\begin{align}
     &\dot{n}_{\vec{q}}^\mathrm{(2)}
    =
    \sum_{\vec{q}'}
    \Big(
    (n_{\vec{q}}(T_\mathrm{m})+1)
    n_{\vec{q}'}(T_\mathrm{m})
     n_{\vec{q}-\vec{q}'}(T_\mathrm{e})
     + 
    (\vec{q}\leftrightarrow \vec{q}')
    \Big)
    \Gamma_{\vec{q}\vec{q}'}
    \\
    &=
    \sum_{\vec{q}'}
    \Big(
    n_{\vec{q}'-\vec{q}}(T_\mathrm{m})
     n_{\vec{q}-\vec{q}'}(T_\mathrm{e}) - 
    (\vec{q}\leftrightarrow \vec{q}')
    \Big)
    \times
     \big(n_{\vec{q}}(T_\mathrm{m}) -n_{\vec{q}'}(T_\mathrm{m})\big)
    \Gamma_{\vec{q}\vec{q}'}
    \\
    &=
    \sum_{\vec{q}'}
    \frac{1}{2}
    \bigg(
    \coth\bigg(\frac{\hbar(\omega_{\vec{q}'}-\omega_{\vec{q}})}{2k_\mathrm{B}T_\mathrm{e}}\bigg)
     -
    \coth\bigg(\frac{\hbar(\omega_{\vec{q}'}-\omega_{\vec{q}})}{2k_\mathrm{B}T_\mathrm{m}}\bigg)
    \bigg)
     \big(n_{\vec{q}}(T_\mathrm{m}) -n_{\vec{q}'}(T_\mathrm{m})\big)
    \Gamma_{\vec{q}\vec{q}'}
    \\
    &\approx
    \sum_{\vec{q}'}
    \frac{n_{\vec{q}}(T_\mathrm{m}) -n_{\vec{q}'}(T_\mathrm{m})}{\hbar(\omega_{\vec{q}'}-\omega_{\vec{q}})}
     k_\mathrm{B} (
    T_\mathrm{e}
     -
    T_\mathrm{m}
    )
    \Gamma_{\vec{q}\vec{q}'}
    \\
    &\approx
    \sum_{\vec{q}'}
    \frac{\partial n_{\vec{q}}(T_\mathrm{m})}{\partial (\hbar \omega_{\vec{q}}) }
     k_\mathrm{B} (
    T_\mathrm{m}
     -
    T_\mathrm{e}
    )
    \Gamma_{\vec{q}\vec{q}'}
    \\
    &=
    \sum_{\vec{q}'}
    \frac{\partial n_{\vec{q}}(T)}{\partial T }\bigg|_{T=T_\mathrm{m}}
     \frac{k_\mathrm{B}T_\mathrm{m} }{\hbar \omega_{\vec{q}}}
     (
    T_\mathrm{e}
     -
    T_\mathrm{m}
    )
    \Gamma_{\vec{q}\vec{q}'}
    \\
    &=
    \sum_{\vec{q}'}
    C_{\vec{q}}
     \frac{k_\mathrm{B}T_\mathrm{m} }{(\hbar \omega_{\vec{q}})^2}
     (
    T_\mathrm{e}
     -
    T_\mathrm{m}
    )
    \Gamma_{\vec{q}\vec{q}'}.
\end{align}
Using the expressions for $\dot{n}_{\vec{q}}^\mathrm{(1)}$ and $\dot{n}_{\vec{q}}^\mathrm{(2)}$, the change in total energy of the magnons can then be calculated as
\begin{align}
    \frac{\partial E_\mathrm{m}}{\partial t}
    &=
    \frac{\partial E_\mathrm{m}}{\partial T_\mathrm{m}}
    \frac{\partial T_\mathrm{m}}{\partial t}
    =
    \underbrace{
    \sum_{\vec{q}} \hbar \omega_{\vec{q}}
    \frac{\partial n_{\vec{q}}(T)}{ \partial T}|_{T=T_\mathrm{m}}
    }_{C_\mathrm{m}}
    \frac{\partial T_\mathrm{m}}{\partial t}=
    (
    T_\mathrm{e}
    -
    T_\mathrm{m}
    )
    \underbrace{
    \sum_{\vec{q}} 
    C_{\vec{q}}
    \Big(
    \gamma_{\vec{q}}
    +
    \sum_{\vec{q}'}
    \frac{k_\mathrm{B}T_\mathrm{m} }{\hbar \omega_{\vec{q}}}
    \Gamma_{\vec{q}\vec{q}'}
    \Big) .
    }_{G_\mathrm{em}} 
\end{align}
With that, the (N+2)TM transforms into the 3TM (in the absence of magnon-phonon coupling), which is given by
\begin{align}
    \begin{split}
    C_\mathrm{m}
      \dot{T}_\mathrm{m}
    &=
  G_\mathrm{em}
      (
    T_\mathrm{e}
    -
    T_\mathrm{m}) ,
    \\
    C_\mathrm{e}
    \dot{T_\mathrm{e}} ~
    &=
    G_\mathrm{em}
    (
    T_\mathrm{m}
    -
    T_\mathrm{e})
    +
    G_\mathrm{ep}(T_\mathrm{p}-T_\mathrm{e})
    +
    P(t) ,
    \\
    C_\mathrm{p}
    \dot{T_\mathrm{p}}
    &
    =
   G_\mathrm{ep}
    (T_\mathrm{e}-T_\mathrm{p}).
    \end{split}
\end{align}

%\newpage

\subsection{\textit{Ab initio} calculations}
\label{AppendixC}

To obtain a full solution of the (N+2)TM, it is necessary to compute the material specific quantities $\Delta$, $\gamma_{\vec{q}}$, $\Gamma_{\vec{q}\vec{q}'}$ and $I_{\sigma\sigma}(T_\mathrm{e})$. For this purpose, we use the full-potential linear augmented plane wave code ELK \cite{ELK}.

As a first step, we determine the coupling parameter $\Delta$ of the $sp-d$ model, which sets the general scale of the electron-magnon scattering. As shown in the main text, the first term (zeroth order in magnon variables) in the electron-magnon scattering Hamiltonian reads $\hat{H}^\mathrm{(0)}_\mathrm{em}= -    \Delta    \sum_{\vec{k}\nu}    (    \hat{c}^\dagger_{\vec{k}\nu\uparrow}\hat{c}_{\vec{k}\nu\uparrow}    -    \hat{c}^\dagger_{\vec{k}\nu\downarrow}\hat{c}_{\vec{k}\nu\downarrow})$, with $\nu \in \{s,p\}$. Based on this, $\Delta$ can be estimated from the projected density of states (DOS), since it is one half of the spin-dependent energy splitting of the $s$- and $p$-bands. In general, this splitting may vary for different electronic states. This is not accounted for in the model used here, where instead a single parameter is used to model the spin splitting. We find, however, that for bcc iron this is justified, since the shift in both $s$- and $p$-bands around the Fermi energy -- the relevant region for electron-magnon scattering -- between spin up and down states is approximately constant with a value of $\Delta\approx \SI{0.75}{\electronvolt}$, see left panel of Figure \ref{fig:pdos}.

Now we calculate the first and second order scattering rates using the formulas derived above,
\begin{align}
    \label{eq:SM_gamma1}
    \gamma_{\vec{q}}
    &=
    \frac{4\pi\Delta^2}{ SN}
    \omega_{\vec{q}}
    I_{\uparrow\downarrow}(T_\mathrm{e})
    \sum_{\vec{k}\nu \nu'}
   \delta( {\varepsilon}_\mathrm{F} - {\varepsilon}_{\vec{k}-\vec{q}\nu \uparrow})
    \delta( {\varepsilon}_\mathrm{F} - {\varepsilon}_{\vec{k}\nu'\downarrow}) ,
    \\
    \begin{split}
    \label{eq:SM_gamma2}
    \Gamma_{\vec{q}\vec{q}'}
    &=
    \frac{2\pi\Delta^2}{S^2N^2}
    ( \omega_{\vec{q}}
    -
    \omega_{\vec{q}'})
    \sum_{\sigma}
     I_{\sigma\sigma}(T_\mathrm{e})
    \sum_{\vec{k}\nu\nu'} 
    \delta(\varepsilon_\mathrm{F}-\varepsilon_{\vec{k}-\vec{q}+\vec{q}'\nu\sigma })
    \delta(\varepsilon_\mathrm{F}-\varepsilon_{\vec{k}\nu'\sigma }).
    \end{split}
\end{align}
The calculation of both quantities requires a spin-dependent summation over the Fermi surface, analogous to what was done in Ref.~\cite{Carva2011} for the evaluation of the spin-dependent Eliashberg function for electron-phonon scattering. As in Ref.~\cite{Carva2011} we use a Gaussian broadening of the Dirac delta distributions by $\SI{0.03}{\electronvolt}$. Also, since we only include the contribution of $s$- and $p$-states (indicated by $\nu,\nu'$) to the scattering, we have to project the Kohn-Sham states (indicated by $n,n'$) onto the spherical harmonics $Y_l^m$ via
\begin{align}
    \delta(\varepsilon_\mathrm{F}-\varepsilon_{\vec{k}\nu\sigma })
    \delta(\varepsilon_\mathrm{F}-\varepsilon_{\vec{k}'\nu'\sigma' })
    &=
    \sum_{nn'}
    P_{\vec{k}\sigma}^{n\nu}
    P_{\vec{k}'\sigma'}^{n'\nu'}
    \delta(\varepsilon_\mathrm{F}-\varepsilon_{\vec{k}n\sigma })
    \delta(\varepsilon_\mathrm{F}-\varepsilon_{\vec{k}'n'\sigma' }),
\end{align}
with $P_{\vec{k}\sigma}^{n\nu}$ being projector functions.

The functions $I_{\sigma\sigma'}(T_\mathrm{e})$ describe corrections {to the scattering rate} at high electron temperatures and are given by
\begin{equation}
    I_{\sigma\sigma'}(T_\mathrm{e})
    =
    \int d{\varepsilon}\
    (-1)
    \frac{\partial f^\mathrm{FD}({\varepsilon},T_\mathrm{e})}{\partial {\varepsilon}}
     \frac{g_\sigma( {\varepsilon})g_\sigma'( {\varepsilon})}{g_\sigma({\varepsilon}_\mathrm{F}) g_\sigma'({\varepsilon}_\mathrm{F})},
\end{equation}
with $g_\sigma(\varepsilon) = \sum_{\vec{k}\nu} \delta(\varepsilon-\varepsilon_{\vec{k}\nu\sigma}) = \sum_{\vec{k}\nu} \sum_{n} P_{\vec{k}\sigma}^{n\nu}\delta(\varepsilon-\varepsilon_{\vec{k}n\sigma})$ being the cumulative DOS of both $s$- and $p$-states. We find that they increase monotonously with the electron temperature (see right panel of Figure \ref{fig:pdos}). However, even for temperature up to $\SI{2000}{\kelvin}$, the {$I_{\sigma\sigma'}(T_{\rm e})$ functions} are below two. Hence, we concluded that the approximation $I_{\sigma\sigma'}=1$ is reasonable for the laser fluences -- heating the electrons up to around $\SI{700}{\kelvin}$ -- considered in the main text.

Figure~\ref{fig:Fig5} depicts the numerically calculated scattering rates using $I_{\sigma\sigma'}=1$ and $\Delta=\SI{0.75}{\electronvolt}$ as obtained above. In the left panel, we show the scattering rate $\gamma_{\bm q}$ that is first order in the magnon variables  {through color code on the magnon dispersion}. It is strictly positive and tends to increase with magnon frequency. The right panel shows  the effective scattering rate $\gamma_{\vec{q}}^\mathrm{(2)}=\sum_{\vec{q}'}\Gamma_{\vec{q}\vec{q}'}$ due to the scattering term second order in magnon variables. Notably, this quantity is negative for low frequencies and positive for high frequencies, indicating that it leads to a depopulation of magnons at low energies due a scattering from low to high energies (the total magnon number is kept constant). In general, the values of the effective second order scattering rate are comparable to the one first order in magnon variables. They are, however, distributed differently: e.g., for magnons close to the $\Gamma$ point the second order scattering rate is by far the dominating one. This is the reason why, as demonstrated in the main text, a laser pulse can in fact lead to a cooling of low energy magnons, i.e., to a decrease of their occupation numbers.

Lastly, we show in Figure \ref{fig:SM_disp_alpha} 
the \textit{ab initio} computed mode-dependent Gilbert damping,
$\alpha_{\bm{q}} = \omega_{\bm{q}}/\gamma_{\bm{q}}$. Interestingly, the Gilbert damping $\alpha_{\bm{q}} $ is large ($\sim 0.01$) at the BZ center and at the high-symmetry points H, N and P at the BZ edge. There is also a noticeable directional anisotropy in the Gilbert damping for modes along $\Gamma -$H and $\Gamma -$P.
We emphasize that the Gilbert damping is here due to the electron-magnon scattering term that is first order in the magnon variables. Other scattering mechanisms as phonon-magnon scattering could contribute further to the mode-specific Gilbert damping.

\begin{figure}[ht!]
    \begin{subfigure}
    \centering
    \includegraphics{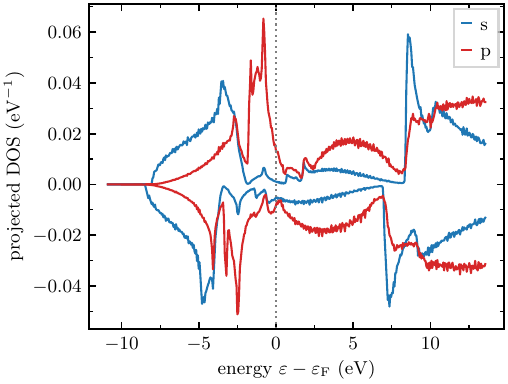}  
    \end{subfigure}
    \begin{subfigure}
    \centering
    \includegraphics{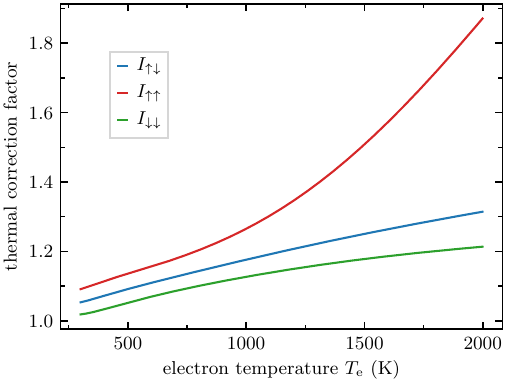}    
    \end{subfigure}
    \caption{\textit{Left}: Projected spin-polarized DOS for bcc iron. Spin-minority density is shown by positive values, spin-majority density by negative values. The exchange splitting is $2\Delta\approx \SI{1.5}{\electronvolt}$ in a large interval around the Fermi energy and for both $s$- and $p$-states. \textit{Right}: Thermal correction factors $I_{\sigma\sigma'}$ versus electron temperature $T_\mathrm{e}$ calculated from the projected DOS.}
    \label{fig:pdos}
\end{figure}

\begin{figure}
    \begin{subfigure}
    \centering
    \includegraphics{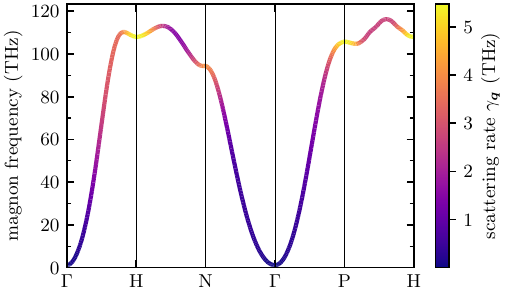}
    \label{fig:SM_disp_1}
    \end{subfigure}
    \begin{subfigure}
    \centering
    \includegraphics{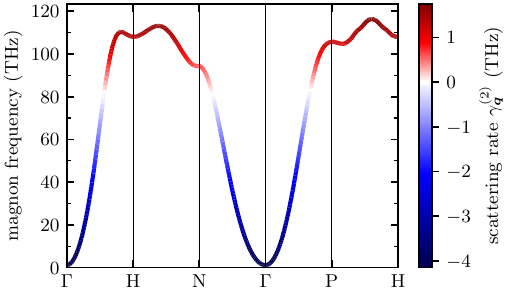}
    \label{fig:SM_disp_2}
    \end{subfigure}
    \caption{
    Magnon dispersion of bcc iron 
    %based on parameters from Ref.~\cite{Mryasov1996} 
    along high-symmetry lines of the Brillouin zone.  The color coding describes (\textit{left}) the scattering rates $\gamma_{\bm{q}}$ due to the electron-magnon scattering term first order in magnon variables $\gamma_{\vec{q}}$ %(\textit{left}) 
    and (\textit{right}) the effective scattering rate $\gamma_{\vec{q}}^\mathrm{(2)}=\sum_{\vec{q}'}\Gamma_{\vec{q}\vec{q}'}$ due to the term second order in magnon variables, calculated with $I_{\sigma\sigma'}=1$ and $\Delta=\SI{0.75}{\electronvolt}$.}
    \label{fig:Fig5}
\end{figure}

\begin{figure}
    \centering
    \includegraphics{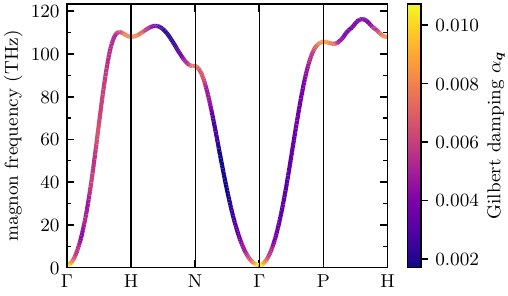}
    \caption{Calculated mode-specific Gilbert damping $\alpha_{\bm{q}} = \omega_{\bm{q}}/\gamma_{\bm{q}}$, depicted by the color code on the magnon dispersion of bcc iron.  The mode-specific Gilbert damping $\alpha_{\bm{q}}$ is due to the electron-magnon scattering term first order in magnon variables.}
    \label{fig:SM_disp_alpha}
\end{figure}

\FloatBarrier
\end{widetext}

\begin{acknowledgments}
The authors thank K.\ Carva for valuable discussions. This work has been supported by the Swedish Research Council (VR), the German Research Foundation (Deutsche Forschungsgemeinschaft) through CRC/TRR 227 ``Ultrafast Spin Dynamics" (project MF, {\blue project-ID: 328545488}), and the K.\ and A.\ Wallenberg Foundation (Grant No.\ 2022.0079). Part of the calculations were enabled by resources provided by  the National Academic Infrastructure for Supercomputing in Sweden (NAISS) at NSC Link\"oping partially funded by the Swedish Research Council through grant agreement No.\ 2022-06725.
\end{acknowledgments}

\section*{Conflict of Interest}
The authors declare no conflict of interest.
\section*{Data Availability Statement}
Data available on request from the authors.

\section*{Keywords}
Ultrafast magnetism, electron-magnon coupling, non-thermal magnons

%\bibliography{bibfile.bib}
%\bibliographystyle{apsrev4-1}
%merlin.mbs apsrev4-1.bst 2010-07-25 4.21a (PWD, AO, DPC) hacked
%Control: key (0)
%Control: author (72) initials jnrlst
%Control: editor formatted (1) identically to author
%Control: production of article title (-1) disabled
%Control: page (0) single
%Control: year (1) truncated
%Control: production of eprint (0) enabled
%

\end{document}